\documentclass[12pt,a4paper]{article}
\usepackage{amsmath,amssymb,mathrsfs,framed,esint}
\usepackage[colorlinks]{hyperref}
\usepackage{color}
\usepackage[all]{xy}

\newcommand{\rem}[1]{}
\newcommand{\bc}{{\boldsymbol{c}}}

\newcommand{\de}{{\rm d}}

\newcommand{\bn}{{\boldsymbol{a}}}

\newcommand{\bX}{{\mathbf{X}}}
\newcommand{\bx}{{\mathbf{x}}}
\newcommand{\bv}{{\mathbf{v}}}
\newcommand{\bw}{{\mathbf{w}}}
\newcommand{\bb}{{\boldsymbol{b}}}
\newcommand{\bA}{{\mathbf{A}}}

\newcommand{\bE}{{\mathbf{E}}}
\newcommand{\bB}{{\mathbf{B}}}

\newcommand{\bu}{{\boldsymbol{u}}}

\newcommand{\bOmega}{{\boldsymbol{\Omega}}}
\newcommand{\bomega}{{\boldsymbol{\omega}}}
\newcommand{\bkappa}{{\boldsymbol{\kappa}}}
\newcommand{\bnu}{{\boldsymbol{\nu}}}

\newcommand{\bgamma}{{\boldsymbol{\gamma}}}
\newcommand{\brho}{{\boldsymbol{\rho}}}

\newcommand{\bmu}{\boldsymbol{\mu}}

\newcommand{\beq}{\begin{equation}}
\newcommand{\eeq}{\end{equation}}
\newcommand{\ben}{\begin{eqnarray}}
\newcommand{\een}{\end{eqnarray}}

\begin{document}

\title{From liquid crystal models to the\\guiding-center theory of magnetized plasmas\\}

\author{Cesare Tronci\\
\it\footnotesize Department of Mathematics, University of Surrey, Guildford GU2 7XH, United Kingdom\\
%\it\footnotesize $^2$Section de Math\'ematiques, \'Ecole Polytechnique
%F\'ed\'erale de Lausanne, Switzerland\vspace{-.2cm}
}

\date{}

\maketitle

\begin{abstract}  
\footnotesize
Upon combining Northrop's picture of charged particle motion with modern liquid crystal theories, this paper provides a new description of guiding center dynamics (to lowest order). This new perspective is based on a  rotation gauge field (\emph{gyrogauge}) that encodes rotations around the  magnetic field. In liquid crystal theory, an analogue rotation field is used to encode the rotational state of rod-like molecules.  Instead of resorting to  sophisticated tools (e.g. Hamiltonian perturbation theory and Lie series expansions) that still remain essential in higher-order gyrokinetics, the present approach combines the WKB method with a simple kinematical ansatz, which is then replaced into the charged particle Lagrangian. The latter is eventually averaged over the gyrophase to produce  Littlejohn's guiding-center equations. A crucial role is played by  the vector potential for the gyrogauge field. A similar vector potential is related to liquid crystal defects and is known as \emph{wryness tensor} in Eringen's micropolar theory. \\

\end{abstract}

{\small
\tableofcontents
}

%--------------------------------------------------------

\section{Introduction}

This paper aims to establish some analogies between liquid crystal models and the  theory of guiding center motion in plasma physics. Evidently, these two theories arose in very different fields of physics and their relationships are far from obvious. However, this paper shows that the gyroradius of Northrop's guiding center picture \cite{Northrop,Northrop1} and the director field of uniaxial liquid crystal molecules (in both the Ericksen-Leslie and Eringen's formulations) possess several similarities that can be exploited to shed new light on Littlejohn's geometric use of gyro-gauge potentials \cite{Littlejohn88}. Indeed, the concept of rotational state of a rod-like nematic molecule can be transferred to define a rotational state of the gyroradius. The essential difference between the two concepts is that the gyroradius can only rotate around the magnetic field vector and this is encoded by using the axis-angle parametrization of three-dimensional rotations (Rodrigues' formula). Despite this difference, the two concepts are related by common geometric features,
\rem{ %%%%%%%%%%%%%%%%%%%%%%%%%%%%%%%%%%%%%%%%
. On one hand, the axis-angle parametrization discloses an essential geometric difference between the two constructions: while the rotational state of nematic molecules is given in terms of a local 3D rotational gauge, rotations of the gyroradius can be identified with global 2D rotations (preserving the magnetic field locally). On the other hand, 
}  %%%%%%%%%%%%%%%%%%%%%%%%%%%%%%%%%%%%%%%%
which can be exploited by the use of rotation matrix fields. In turn, this leads in both cases to the emergence of a certain gauge potential that occurs in Eringen's micropolar theory \cite{Eringen1997} and is a key geometric object in Littlejohn's theory of guiding center motion \cite{Littlejohn1981,Littlejohn1982,Littlejohn88,Littlejohn}.

These analogies can be pushed even further by considering biaxial nematic molecules (as described by a pair of orthogonal directors) and the pair of unit vectors associated to the magnetic field and the gyroradius. This analogy is particularly transparent in the case of time-dependent magnetic field, although guiding center theory requires splitting rotations around the magnetic field from rotations {of} the magnetic field itself and this splitting introduces an composition of two- and three-dimensional rotations that leads to an interesting form of the gauge potential. 

In order to provide a systematic exposition, we shall first review the essential concepts associated to the rotational state of liquid crystal molecules. Then, this introduction will proceed to illustrate the main ideas underlying Northrop's guiding center picture. The main analogies will also be discussed.

\subsection{Uniaxial nematic phases and the director field\label{nematics}}

Liquid crystals are well known examples of fluids with internal (micro-)structure \cite{Chandra1992,KlLa2003}. More particularly, liquid crystals are typically modeled as fluids carrying particles that are endowed with an orientational state. This orientational state is incorporated in an additional microscopic variable, that is known in condensed matter theory as \emph{order parameter}. This order parameter emerges when the full rotational symmetry is broken by the particular shape of the particle (or molecule).  Several types of liquid crystal phases are available and each of them is modeled by a different type of order parameter in some coset space \cite{Mermin1979,Mi1980}. In the simplest case of nematic liquid crystals, rod-like molecules carry a preferred direction in space. Then, the order parameter is identified by an unsigned unit vector $\bf n$, called the \emph{director}. In the more complicated case of biaxial nematic molecules, the order parameter is identified with a pair of orthogonal directors. 

When one aims to consider the entire liquid crystal flow, one builds a continuum theory by replacing the order parameter of the single molecule by an order parameter field, which is evaluated at the Lagrangian particle position.
The most celebrated theory of uniaxial nematic flows was formulated by Ericksen and Leslie \cite{Le1979}, who coupled Euler's fluid equation to an Euler-Lagrange equation for the director field $\mathbf{n(x},t)$. The same theory can be expressed in terms of the director field $\mathbf{n(x},t)$ and its angular velocity field $\boldsymbol{\omega}$ such that $\partial_t{\mathbf{n}}=\boldsymbol{\omega}\times\mathbf{n}$. One way the angular velocity is naturally incorporated in the treatment is by writing the director dynamics in terms of the rotational state $\mathcal{R}$ as follows:
\beq\label{director}
\mathbf{n(x},t)=\mathcal{R}({\bf x},t)\mathbf{n}_0(\mathbf{x})
\,,
\eeq
where $\mathcal{R}$ is a time-dependent rotation gauge field, so that $\mathcal{R}({\bf x},t)$ is a rotation matrix at all times and at all points in space (such that $\mathcal{R}({\bf x},0)=\boldsymbol{1}$). Here, we have not considered the fluid motion as this is not essential for our purposes. This approach was followed by Dzyaloshinskii and Volovik \cite{Volovick1980} to formulate a Poisson-bracket structure to uniaxial nematic flows. The same idea was later implemented in the Lagrangian variational framework \cite{{GBTr2010},Ho2002}. In the latter case, the angular velocity appears under its skew-symmetric matrix correspondent $\widehat{\omega}$ as follows:
\beq
\partial_t\mathbf{n}=(\partial_t\mathcal{R}) \mathcal{R}^{-1}\mathbf{n}=\widehat{\omega}\mathbf{n}
\,,
\label{directorfrequency}
\eeq
so that $\widehat{\omega}=(\partial_t\mathcal{R}) \mathcal{R}^{-1}$ and the angular velocity vector is given in components by $\omega_a=\varepsilon_{abc}\widehat{\omega}_{cb}/2$. 

Then, the rotation field becomes a local gauge field and the whole liquid crystal theory can be formulated as a gauge theory. For example, a gauge vector potential emerges upon writing 
\[
\nabla_i\mathbf{n}=(\nabla_i\mathcal{R}) \mathcal{R}^{-1}\mathbf{n}-\mathcal{R}\widehat{\gamma}_{0\,i}\mathcal{R}^{-1}\mathbf{n}
\,,
\]
where $\nabla_i$ denotes the $i$th spatial derivative and $\widehat{\gamma}_0=\widehat{\gamma}_{0\,i}^{\; a}\,\de x^i\,\mathbf{e}_a$ corresponds to a tensor such that $\nabla_i\mathbf{n}_0=-{\bgamma}_{0\,i}\times\mathbf{n}_0$ (for example, ${\bgamma}_{0\,i}=-\mathbf{n}_0\times\nabla_i\mathbf{n}_0$). Eventually, one is left with the relation 
\[
\nabla_i\mathbf{n}=-\bgamma_i\times\mathbf{n}
\,,
\]
and the definition of the gauge potential
\beq\label{wryness}
\widehat{\gamma}_i=\mathcal{R}\nabla_i\mathcal{R}^{-1}+\mathcal{R}\widehat{\gamma}_{0\,i}\mathcal{R}^{-1}
\,,
\eeq
where we notice that $\widehat{\gamma}(\bx,t=0)=\widehat{\gamma}_0$. This evolution law produces the equation of motion 
\beq
\partial_t\bgamma_i=-\nabla_i\bomega-\bgamma_i\times\bomega
\,. 
\label{gamma-eq1}
\eeq
In Eringen's micropolar theory \cite{Eringen1997}, the tensor \eqref{gamma-eq1} is called \emph{wryness tensor} and quantifies the spatial rotational strain, that is the amount 
by which a specified director field rotates under an infinitesimal 
displacement \cite{Ho2002}. Its gauge theory interpretation was recently exploited in \cite{GBRa2009,GBRaTr2012,GBRaTr2011} to show the equivalence of the Ericksen-Leslie model with Eringen's theory. Notice that $\bgamma$ cannot be expressed in terms of the director, since the parallel component $\bgamma_i\cdot{\bf n}$ is undetermined. On the other end, in the case of biaxial molecules, the existence of two orthogonal directors $({\bf n,m})$ is equivalent to identifying the order parameter with the whole rotation matrix $\mathcal{R}$ and therefore $\bgamma$ can be indeed expressed in terms of the directors. Indeed, if we define a ${\bgamma}_{0}$ such that $({\bf \nabla n}_0,{\bf \nabla m}_0)=-(\bgamma_0\times{\bf n}_0,\bgamma_0\times{\bf m}_0)$, then the wryness tensor \eqref{wryness} satisfying $({\bf \nabla n},{\bf \nabla m})=-(\bgamma\times{\bf n},\bgamma\times{\bf m})$ at all times is expressed equivalently as (analogous formulas appeared in \cite{Ho2002})
\begin{align}
\bgamma_i&=\nabla_i{\bf n}\times{\bf n}+\left(\nabla_i{\bf m}\cdot{\bf m}\times{\bf n}\right){\bf n}
\nonumber
\\
&=\nabla_i{\bf m}\times{\bf m}+\left(\nabla_i{\bf n}\cdot{\bf n}\times{\bf m}\right){\bf m}
\,.
\label{wrynessbiaxial}
\end{align}

This non-Abelian gauge-theory approach   applies to very general systems since it incorporates defect dynamics in various
models, such as frustrated spin glasses 
\cite{Volovick1980,HoKu1988}. 
In the gauge theoretical 
setting, one interprets the wryness tensor $\boldsymbol{\gamma}$ as 
the magnetic potential of a static Yang-Mills magnetic field. Then, its corresponding magnetic  field  is given  by
$
\boldsymbol{\cal B}^i=\epsilon^{ijk}\!\left(\partial_j\bgamma_k+\bgamma_j\times\bgamma_k\right)$,
where summation over repeated indices is understood. Orientational defects are identified with points (or sets of points) where the director field is not defined.  In this gauge theory approach
(see \cite{Volovick1980})  a vanishing
 Yang-Mills magnetic field $\boldsymbol{\cal B}$  characterizes the absence of   orientational defects, which are instead allowed when $\boldsymbol{\cal B}\neq 0$.

As it was shown extensively by Littlejohn \cite{Littlejohn}, an alternative gauge theory construction is also of fundamental importance in the geometry of guiding center motion, although its analogies with liquid crystal dynamics have never been disclosed. More particularly, upon identifying $\bf n$ with the unit gyroradius and $\bf m$ with the unit magnetic field vector, analogous formulas emerge in guiding center theory and the component of $\bgamma_i$ along the gyroradius direction equals (up to a sign) a phase-type gauge potential, typically denoted by $\bf R$ in Littlejohn's work \cite{Littlejohn,Littlejohn88}. These features are presented explicitly in the remainder of this paper. The next Section introduces Northrop's picture of guiding center motion \cite{Northrop,Northrop1} and introduces the general setting for this work.

\subsection{Lorentz force and guiding center motion}

Guiding center theory was first proposed by Alfv\'en \cite{AlFa} and further developed by Northrop \cite{Northrop,Northrop1}, as an attempt to average out the fast gyro-motion of a charged particle around a strong magnetic field. These early formulations suffered from several drawbacks, mainly due to the fact that they did not conserve energy in the general case \cite{CaBr}. The final theory of guiding center motion was then formulated by Littlejohn \cite{Littlejohn1981,Littlejohn}, who used sophisticated techniques such as Hamiltonian perturbation theory for Poisson brackets and Lie series expansions of the Poincar\'e-Cartan form \cite{CaLi,Littlejohn1981,Littlejohn1982,Littlejohn1993}. These deeply geometric tools are widely recognized as necessary for a self-consistent guiding-center theory, which is also the basis of modern gyrokinetic models \cite{BrHa,Burby}. This paper aims to present an alternative approach that is based on analogies with liquid crystal dynamics. As we shall see, this approach is used in the lowest order-theory of guiding center motion, while higher-order augmentations still require geometric perturbation methods for Hamiltonian structures \cite{CaLi,Littlejohn1981}.

Upon starting with the Lorentz force equation 
\beq\label{Lorentz}
m\ddot{\bf r}=q\big(\bE({\bf r},t)+\dot{\bf r}\times\bB({\bf r},t)\big)
\eeq
for a point charge in an external electromagnetic field, 
the fundamental idea of the guiding center approximation is that the charge position ${\bf r}(t)$  can be expanded in \eqref{Lorentz} as
\beq\label{gyroexpansion}
{\bf r}(t)=\bX(t)+\varepsilon\brho(t)
\,,
\eeq
where the \emph{gyroradius} $\brho(t)$ is a vector gyrating on the plane orthogonal to the magnetic field $\bB(\bX(t),t)$  evaluated at $\bX(t)$ (so that $\brho(t)\cdot\bB(\bX(t),t)=0$). The vector $\bX(t)$ is the position of the \emph{guiding center}, which carries the information about the motion along the field direction. The small parameter $\varepsilon$ deserves some attention. As Northrop \cite{Northrop,Northrop1} showed, working with dimensionless quantities in \eqref{Lorentz} leads to identifying $\varepsilon$ with the ratio of the Larmor gyroradius to the characteristic distance over which fields change. However, Northrop also noticed that  this is equivalent to 
identifying $\varepsilon$ with the dimensional charge-to-mass ratio (that is $\varepsilon=q/m$) and this avoids the necessity of rescaling the Lorentz-force equation \eqref{Lorentz}. We notice that this procedure prevents the variable $\brho$ in \eqref{gyroexpansion} from possessing the dimension of a length. Therefore, we should bear in mind that giving up on nondimensionalization means that the physical Larmor gyroradius is actually 
\beq\label{gyroradiusconversion}
\brho_L=\varepsilon\brho
\eeq
and not $\brho$ itself. Nevertheless, we shall keep calling $\brho$ the `gyroradius' in order to simplify the discussion and we shall consider the relation \eqref{gyroradiusconversion} for later use. Dimensional arguments of this kind will apply to the remainder of this work, while they would be unnecessary if we were working with dimensionless quantities.

Northrop's approach continues by replacing \eqref{gyroexpansion} in the Lorentz force equation, which is then expanded to first order. As a further step, one adds the information that $\brho$ gyrates in the plane orthogonal to $\bB(\bX(t),t)$ by inserting the following expression for the gyroradius:
\beq\label{Northrop}
\brho(t)=\rho(t)\left(\mathbf{e}_1\cos\vartheta(t)-\mathbf{e}_2\sin\vartheta(t)\right)
\,,\qquad\quad 
\dot{\vartheta}(t)=\varepsilon^{-1}\bB(\bX(t),t)
\,,
\eeq
where $\mathbf{e}_1(\bx,t)$ and $\mathbf{e}_2(\bx,t)$ are orthogonal unit vectors  such that $\bb(\bx,t)={\mathbf{e}_1(\bx,t)\times\mathbf{e}_2(\bx,t)}$ and the whole triad is evaluated at the guiding center position. Here, we have used the notation $\bB= B\bb$  along with Littlejohn's conventions \cite{Littlejohn} in defining $\mathbf{e}_1$ and $\mathbf{e}_2$. As a further step, averaging over the \emph{gyrophase} $\vartheta$ yields Northrop's \emph{basic equation of guiding center motion} (see (1.12) in \cite{Northrop,Northrop1})
\beq\label{NGC}
m\ddot\bX=q\big(\bE(\bX,t)+\dot\bX\times\bB(\bX,t)\big)-\mu\nabla B(\bX,t)
\,.
\eeq
where $\mu%=\frac{q^2}{2m}\rho_L^2(t)B(\bX(t))
=-({m}/2)\rho^2(t)B(\bX(t))$ is the magnitude of the magnetic moment. 
In his approach, Northrop proceeds by constructing the phase-space picture of the equation above and by making further approximations, as they arise from the assumption that the fields vary on lengthscales much bigger than the Larmor gyroradius \eqref{gyroradiusconversion}. As mentioned previously, the resulting equations do not conserve energy in the general case \cite{CaBr}. This lack of energy balance is related to the fact that Northrop's equations on phase-space do not possess a Hamiltonian structure, which was instead recovered in Littlejohn's theory of guiding center motion \cite{Littlejohn}. Littlejohn's basic equation of motion reads (see equation (6) in \cite{Littlejohn})
\beq\label{LGC}
\bE-\varepsilon u\partial_t\bb-\mu\nabla B+\varepsilon^{-1}\dot{\bX}\times(\bB+\varepsilon u \nabla\times\bb)=\dot{u}\bb
\,,\qquad\quad 
u=\bb\cdot\dot\bX\,,
\eeq
where $\bE=-\partial_t\bA-\nabla\phi$ and all field variables are evaluated at the guiding center position $\bX$. Then, crossing with $\bb$ and dotting with $\bB^*=\bB+\varepsilon u \nabla\times\bb$ yields the phase-space equations
\beq
\dot{\bX}=\frac{u\bB^*+\varepsilon\bE^*\times\bb}{\bb\cdot\bB^*}\,,
\,\qquad\quad
\dot{u}=\frac{\bE^*\cdot\bB^*}{{\bb\cdot\bB^*}}\,,
\label{GCEQ}
\eeq
where $\bE^*=\bE-\varepsilon u\partial_t\bb-\mu\nabla B$. 
These equations conserve energy exactly and, more importantly, they possess a Hamiltonian structure that has been widely studied \cite{CaBr,Littlejohn1981}.

\subsection{Outlook and results}
Although Lie transform techniques and Hamiltonian perturbation theory have been widely successful in the formulation of self-consistent guiding center theory at all orders, this work shows an alternative construction that applies to lowest order. Without requiring any previous knowledge on the gyroradius and the gyrophase (except their fast rotational motion), mathematical techniques in liquid crystal theory are applied in this paper to present an alternative variational construction for guiding center motion. 

More particularly, certain similarities between Northrop's definition \eqref{Northrop} of gyroradius and the director field \eqref{director} of nematic liquid crystals are exploited to show how both Northrop's equation \eqref{NGC} and Littlejohn's theory \eqref{LGC} can be both recovered from simple variational arguments. The approach is based on a combination of the WKB method and the ansatz of rotational dynamics for the gyroradius. Eventually, averaging the Lagrangian yields a gyrophase symmetry that recovers conservation of the magnetic moment.

The main results are 1) the identification of new variational structures for guiding center dynamics (Section \ref{StatMagnF}) and 2) the emergence of a Weyl-type gauge as the natural gauge for guiding center motion in a time-depedent magnetic field (Section \ref{DynMagnF}). The second result is achieved by a separation of slow rotations of the magnetic field and fast rotations around the magnetic field. The introduction of two rotational gauge fields leads to an interesting interplay of fast and slow rotations that lead to new forms of the corresponding gauge potentials.

The discussion starts in the next Section by focusing on the case of static magnetic fields.

\section{Static magnetic fields\label{StatMagnF}}

In this section, we shall show how the gauge theory approach to liquid crystal dynamics applies to guiding center motion, thereby unfolding the variational structure of guiding center motion in both Northrop and Littlejohn pictures. As we shall see, the main advantage is that no previous knowledge of the expressions  of the gyroradius or of the gyrofrequency is required. For example, Northrop approach  assumes an exact expression for the gyrofrequency in \eqref{Northrop}, while Littlejohn's variational approach \cite{Littlejohn} assumes previous knowledge on the expression of gyroradius as it arises from Hamiltonian perturbation theory \cite{Littlejohn1981,Littlejohn1982}. This assumptions are absent in the present approach, which relies entirely on combining a WKB ordering method with a rotational ansatz for the gyroradius. Eventually, averaging the Lagrangian returns a gyrophase symmetry, which produces conservation of the magnetic moment. As its title says, this Section treats the simple case of a static magnetic field, while the dynamical case is treated later.

\subsection{The gyrodirector field\label{sec:gyrodirector}}

This Section focuses on the simple case of a static magnetic field to introduce the analogy between the gyroradius in guiding center theory and the director field  \eqref{director} in nematic media. 
The main concept lies in the definition of a \emph{gyrodirector field} $\bn({\bf x},t)$ (of unit norm) such that the physical gyroradius $\brho(t)$ in \eqref{gyroexpansion} is given as 
\beq\label{gyroradius1}
\brho(t)=\rho(t)\bn(\bX(t),t)
\,.
\eeq
We recall that, in Northrop's approach, equation \eqref{Northrop} yields
\[
\bn(\bX(t),t)=\mathbf{e}_1(\bX(t))\cos\vartheta(t)+\mathbf{e}_1(\bX(t))\times\bb(\bX(t))\sin\vartheta(t)
\,,
\]
which is equivalent to saying that $\bn(\bX(t),t)$ rotates in a plane orthogonal to $\bb(\bX(t))$.

In order to pursue the analogy with the director field, we wish to write the gyrodirector field evolution in an analogous way to \eqref{director}, that is
\beq\label{gyrodirectorevolution}
\bn(\bx,t)=\mathcal{R}(\bx,t)\mathbf{e}_1(\bx)
\,.
\eeq
This relation incorporates the fact that the gyroradius direction evolves by rotating the field $\mathbf{e}_1(\bx)$. However, one still needs to incorporate the fact that the only rotations allowed are those around the magnetic field direction $\bb(\bx)$. For this purpose, we recall the axis-angle parametrization of rotation matrices (Rodrigues' formula): rotating (anti-clockwise) a vector $\bv$ by an angle $\theta$ around the unit vector $\bu$ yields 
\beq\label{Rodrigues}
\mathbf{w}=\bv+\sin\theta\,\bu\times\bv+(1-\cos\theta)\bu\times\bu\times\bv
\,.
\eeq
Then, upon using the hat notation from Section \ref{nematics}, the corresponding rotation matrix $\mathcal{R}(\theta,\bu)$ (function of $\theta$ and $\bu$) such that $\bw=\mathcal{R}(\theta,\bu)\bv$ is written as
\beq\label{Rodrigues2}
\mathcal{R}(\theta,\bu)=:\boldsymbol{1}+\sin\theta\,\widehat{u}+(1-\cos\theta)\,\widehat{u}\,\widehat{u}=e^{\theta \widehat{u}}
\,,
\eeq
where the second equality  follows by expanding out the matrix exponential. 
This leads to the following evolution law for the gyrodirector field
\beq\label{gyrodirector}
\bn(\bx,t)=\mathcal{R}(\theta(t),\bb(\bx))\,\mathbf{e}_1(\bx)
%=e^{\theta(t)\widehat{b}(\bx)}\mathbf{e}_1(\bx)
\,,
\eeq
where we have changed the gyrophase notation because we expect a change in sign, due to the anti-clockwise rotation convention in Rodrigues formula \eqref{Rodrigues}. Multiplying by $\rho(t)$ on both sides and replacing $\theta=-\vartheta$, equation \eqref{gyrodirector} yields the first in \eqref{Northrop}  (see also equations (1.2) in \cite{Littlejohn}), although here we are restricting to consider only static magnetic fields. At the initial time, we set $\theta(0)=0$ so that $\bn(\bx,0)=\mathbf{e}_1(\bx)$ and this is analogous to setting $\vartheta=0$ in equation \eqref{Northrop}, so that $\brho(0)=\rho(0)\,\mathbf{e}_1(\bX(0))$.
At this point, all the relations occurring in Section \ref{nematics} for director evolution can be easily specialized to this case. Notice, however, that in the present case the magnetic field is static and therefore the only dynamical variable in the gyrodirector dynamics \eqref{gyrodirector} is the phase $\theta(t)$ and \emph{not} a whole rotation matrix (as in the case of \eqref{director}). This  actually leads to substantial simplifications that will be partially lost in the case (treated later on) of a dynamic magnetic field.

Upon mimicking the relation \eqref{directorfrequency}, the evolution law \eqref{gyrodirector} yields the gyrodirector frequency
\[
\widehat{\omega}(\bx,t)=(\partial_t\mathcal{R})\mathcal{R}^{-1}=(\partial_te^{\theta(t)\widehat{b}(\bx)})e^{-\theta(t)\widehat{b}(\bx)}=\dot\theta(t)\widehat{b}(\bx)
\,,
\] 
so that, in vector form, the gyrofrequency reads $\bomega(\bx,t)=\dot\theta(t){\bb}(\bx)$. We notice that this relation is greatly simplified in comparison to the case of a dynamic magnetic field. The equation of motion \eqref{directorfrequency} transfers to $\partial_t\bn=\dot\theta\bb\times\bn$. As anticipated earlier, the analogy with director dynamics proceeds further by the identification of a wryness tensor for the gyrodirector. Indeed, if we define $\bgamma_{0\,i}$ such that 
%$\nabla_i\bb=-\bgamma_{0\,i}\times\bb$ and 
$\nabla_i\mathbf{e}_1=-\bgamma_{0\,i}\times\mathbf{e}_1$, one computes
\beq\label{connection1}
\nabla_i\bn=-\bgamma_{i}\times\bn
\,,
\eeq
where $\bgamma$ is given as in \eqref{wryness}. The explicit expression of $\bgamma$ as it arises from \eqref{wryness} is however cumbersome and we shall proceed in a different way. First, we use the fact that the rotation $\mathcal{R}(\theta(t),\bb(\bx))$ preserves $\bb(\bx)$, so that $\nabla(\mathcal{R}\bb)=\nabla\bb$ (in the notation used in Section \ref{nematics} we would write $\bb_0=\mathcal{R}^{-1}\bb=\bb$, where $\bb_0$ is thought of as some `initial' magnetic director). Then, in analogy with the case of biaxial nematics, we recall that $(\bn,\bb)$ is a pair of two orthogonal unit vector and we ask for $\bgamma_{0\,i}$ to satisfy $(\nabla_i\mathbf{e}_1,\nabla_i\bb)=-(\bgamma_{0\,i}\times\mathbf{e}_1,\bgamma_{0\,i}\times\bb)$. Notice that the analogy with biaxial nematics is  even more appropriate if $\bb$ depends on time, although this is not essential. At this point,  we have
\[
\nabla_i\bb=-\bgamma_{0\,i}\times\bb=-\bgamma_i\times\bb
\,,
\]
where the last equality follows by expanding $\nabla(\mathcal{R}\bb)$. Therefore, since $({\nabla_i \bn},{\nabla_i \bb})=-(\bgamma_i\times{\bn}, \bgamma_i\times{\bb})$, proceeding analogously to Section \ref{nematics} yields equation \eqref{wrynessbiaxial} in the form
\[
\bgamma_i=\nabla_i{\bb}\times{\bb}+\left(\nabla_i{\bn}\cdot{\bn}\times{\bb}\right){\bb}
\,,
\]
where we have recalled \eqref{wrynessbiaxial}. Remarkably, since the magnetic field is static and $\nabla_i\bb\cdot\bb=\nabla_i\bn\cdot\bn=0$, one can use vector algebra to compute
$\partial_t\big(\nabla_i{\bn}\cdot{\bn}\times{\bb}\big)=0$ so that
\[
\bb\cdot\bgamma_i=\bb\cdot\bgamma_{0\,i}=-(\nabla_i{\mathbf{e}_1})\cdot{\mathbf{e}_2}
\,,
\]
where ${\mathbf{e}_2}=\bb\times{\mathbf{e}_1}$. Up to a change in sign, this equals precisely Littlejohn's gauge potential 
\[
{\bf R}:=\nabla{\mathbf{e}_1}\cdot{\mathbf{e}_2}
\] appearing in higher-order guiding center theory. This quantity is also related to the anholonomy features in guiding center theory \cite{Littlejohn88}, although this perspective will not be developed in the present work. The vector ${\bf R}$ will reappear later in the case of dynamic magnetic fields. 
Eventually, the wryness tensor of guiding center theory (which we shall call \emph{gyro-wryness tensor}) reads
\[
\bgamma_i(\bx)=\nabla_i{\bb}(\bx)\times{\bb}(\bx)-{R}_i(\bx)\bb(\bx)
\,,
\]
which is constant in time and independent of the gyrophase $\theta$. This conservation of the gyro-wryness tensor is due to the fact that we are considering a static magnetic field and it stands as a substantial difference from Eringen's micropolar theory, where the wryness tensor evolves according to \eqref{gamma-eq1}.
In the case under consideration, the latter simply becomes $\nabla_i\bomega=-\bgamma_i\times\bomega$, which is an obvious consequence of the relations $\nabla_i\bb=-\bgamma_i\times\bb$ and $\bomega=\dot\theta\bb$. As a further remark, we observe that analogously to relations \eqref{wrynessbiaxial}, the gyro-wryness tensor can also be expressed in the equivalent form
$\bgamma_i={\nabla_i{\bn}\times{\bn}}+\left(\nabla_i{\bb}\cdot{\bb}\times{\bn}\right){\bn}$, which can be verified to be constant in time.

Once the geometric kinematic relations are established for the gyrodirector field and its gyro-wryness tensor, the kinematics of the physical gyroradius is completely characterized. Indeed, combining the definition \eqref{gyroradius1} with the relations above yields
\beq\label{gyroradius2}
\brho(t)=\rho(t)\mathcal{R}\big(\theta(t),\bb(\bX(t))\big)\mathbf{e}_{1\!}(\bX(t))
\,.
\eeq
Also, upon taking the phase-average $\langle \cdot \rangle =(1/2\pi) \int_{0}^{2\pi}\!\cdot\,\de \theta$ of the gyroradius, one obtains Northrop's result 
\beq\label{averages}
\langle\brho\rangle=\langle\dot{\brho}\rangle=0
\,,\qquad\quad
\langle\boldsymbol\rho\times\dot{\boldsymbol\rho}\rangle
=\dot{\theta}\rho^2\bb(\bX)
\,,
\eeq
where the latter can be easily written in terms of the magnetic moment vector (recall equation \eqref{gyroradiusconversion})
\beq\label{magneticmoment}
\bmu%=\frac{q^2}{2m}\rho_L^2(t)\bB(\bX(t))
=-\frac{m}2\rho^2(t)\bB(\bX(t))
\,.
\eeq

As we shall see, combining the evolution law \eqref{gyrodirectorevolution} with a first-order expansion and standard averaging returns both Northrop's basic equation \eqref{NGC} and Littlejohn's equations of motion on phase-space \eqref{LGC}, depending on whether one operates in physical space or in phase-space respectively. At the present stage of development, the choice still remains whether one aims to operate on the equations of motion or on the variational principle. These two possible routes may lead to different conclusions and sometimes energy conservation may be even lost if one operates only on the equations of motion (this is the case for Northrop's phase-space theory, as shown in \cite{CaBr}). Therefore,  also in analogy with the recent work on variational principles for liquid crystal dynamics \cite{GBRaTr2012,GBRaTr2011,GBTr2010,Ho2002}, here we shall choose the second option.

\subsection{Magnetic field with constant direction}

In this Section, we shall consider guiding center motion in a magnetic field with constant direction, such that $\nabla\bb=0$. This allows also a convenient choice of the orthogonal vector $\mathbf{e}_1$ in such a way that $\nabla\mathbf{e}_1=0$. This is a great simplification, since the whole gyro-wryness tensor vanishes (i.e. $\bgamma=0$) and the gyrodirector becomes spatially constant ($\nabla\bn=0$). Consequently, the gyroradius and its time-derivative are independent of the guiding centre position and they are written as
\beq
\brho(t)=\rho(t)\bn(t)
\,,\qquad\ 
\dot{\brho}(t)=\rho(t)\dot\theta(t)\bb\times\bn(t)+\dot\rho(t)\bn(t)
\,.
\label{gyrorad-exp}
\eeq
In what follows, we shall apply the kinematic relations discussed so far in two different ways. One is Northrop's picture: the expansion around $\varepsilon$ and the averaging is performed on the particle Lagrangian on the configuration space. The other one is Littlejohn's picture: the expansion around $\varepsilon$ and the averaging is performed on the particle Lagrangian on the configuration space.

\subsubsection{Northrop's picture}
In order to derive guiding center theory from the previous arguments, we replace  \eqref{gyroexpansion} and \eqref{gyroradius2} in the single particle Lagrangian (suitably divided by the particle charge $q$)
\beq\label{Lagr-config}
L({\bf r},\dot{\bf r})=\frac\varepsilon2 \|\dot{\bf r}\|^2+\dot{\bf r}\cdot\bA({\bf r})
\eeq
and expand in terms of $\varepsilon$. 
Notice that in this paper we shall adopt Weyl's gauge $\phi=0$ for the electrostatic potential, although the whole treatment can be easily extended to arbitrary gauges.
In order to keep the correct ordering (recall the second relation in \eqref{Northrop}), we apply the  WKB method by performing the change of variable
\beq
\theta(t)=\varepsilon^{-1}\Theta(t)
\,.
\label{newangle}
\eeq
This relation can be regarded as the lowest order truncation of the WKB-type approximation $\bn=\exp\!\big({\varepsilon^{-1\!}\sum\varepsilon^n\Theta_n \widehat{b}}\,\big)\bn_0$ for the gyrodirector dynamics (upon defining $\Theta=\Theta_1$).
At this point, expanding the particle Lagrangian to first order  (see Appendix \ref{NorthApp}) produces
\beq\label{NorthLagr}
L=\frac\varepsilon2\|\dot\bX\|^2+\dot\bX\cdot\bA(\bX)+ \frac\varepsilon2\|\bOmega\times\brho\|^2
-\frac{\varepsilon}{m}\bmu\cdot\bOmega
%+\frac\varepsilon2\rho^2B\dot\Theta
+\varepsilon\dot\bX\cdot\left( \bOmega+\bB(\bX)\right)\times\brho
\,,
\eeq
where we have defined $\bOmega=\dot\Theta\bb$ and we recall the functional dependence $\brho=\brho(\rho,\Theta)$ and $\bmu=\bmu(\rho,\bX)$. Then, upon dividing by $\varepsilon$, averaging the Lagrangian by using $\langle\brho\rangle=\langle\dot\brho\rangle=0$ yields
\beq\label{NLagrangian}
\langle L\rangle=\frac12\|\dot\bX\|^2+\varepsilon^{-1}\dot\bX\cdot\bA(\bX)+ \frac12\rho_L^2{\dot\theta}\big({\dot\theta}
+ \varepsilon^{-1}B(\bX)\big)
\eeq
where we have restored the variable $\theta=\varepsilon^{-1}\Theta$ and we have used \eqref{gyroradiusconversion}. 
Eventually, taking variations in $\theta$ gives $\rho_L^2B(\bX)=const.$ and variations in $\rho_L$ produce $\dot\theta=-\varepsilon^{-1} B(\bX)$. Although the first relation is the well known invariance of the magnetic moment vector \eqref{magneticmoment}, the second relation differs  from the second in \eqref{Northrop} by a sign. This comes as no surprise, since Rodrigues' formula \eqref{Rodrigues2} applies to counter-clockwise rotations, while a positive charge rotates clockwise. Then, variations in $\bX$ return Northrop's basic equation \eqref{NGC} of guiding center motion in the absence of an electric field, i.e. $\bE=-\partial_t\bA=0$. The degenerate Lagrangian \eqref{NLagrangian} is obtained here for the first time by exploiting the analogies between liquid crystal dynamics and guiding center theory. We shall call \eqref{NLagrangian} 
the \emph{Northrop Lagrangian}. 
At this point, Northrop's analysis continues by writing \eqref{NGC} on phase-space and by making approximations based on the main assumption that the magnetic field varies on lengthscales much bigger than the Larmor gyroradius \eqref{gyroradiusconversion}. These approximations break the variational structure of the equations of motion. In order to avoid this problem, one can follow Littlejohn's approach and apply the approximations on phase space by starting directly with a phase-space Lagrangian \cite{Littlejohn,YeKa}.  Other types of Lagrangians \cite{YeMo} could also be considered.

\subsubsection{Littlejohn's picture}

Littlejohn's picture differs from Northrop's in that it considers phase space dynamics from the beginning. In other words, instead of considering the Lagrangian \eqref{Lagr-config}, Littlejohn considered its phase space counterpart
\beq\label{phspLagr}
L=\big(\varepsilon\bv+\bA({\bf r})\big)\cdot\dot{\bf r}-\frac{\varepsilon}2\|\bv\|^2
\,.
\eeq
In this Section, we shall follow the same steps as in Northrop's picture, though by operating on the phase-space Lagrangian. Then, replacing \eqref{gyroexpansion} and \eqref{gyroradius2} and expanding to first order yields (see Appendix \ref{LittleApp})
\beq
L=(\bv+\varepsilon^{-1}\bA(\bX))\cdot\dot{\bf X}-\frac1m\bmu\cdot\bOmega+\brho\cdot(\dot\bX\times\bB(\bX)+{\bv}\times\bOmega)
%\frac12\rho^2 B\dot\Theta
-\frac{1}2\|\bv\|^2
\,,
\label{LittleLagr}
\eeq
where we have divided by $\varepsilon$. Again, we recall the definition $\bOmega=\dot\Theta\bb$ as well as the functional dependence $\brho=\brho(\rho,\Theta)$ and $\bmu=\bmu(\rho,\bX)$.

At this point, the velocity must be expressed in a convenient frame. Rather than using the fixed frame $(\bb,\mathbf{e}_1,\mathbf{e}_2)$, we shall use the moving frame $(\bn(t),\bb,\bc(t))$ with $\bc(t)=\bn(t)\times\bb=-\mathcal{R}(\theta(t),\bb)\,\mathbf{e}_2$. Therefore we write
\[
\bv(t)
%=u(t)\bb+\bv_\perp(t)
=u(t)\bb+n(t)\boldsymbol{a}(t)+w(t)\boldsymbol{c}(t)\,,
\]
so that %$\bv_\perp(t)=s(t)\boldsymbol{a}(t)+w(t)\bn(t)\times\bb$ and
\[
\brho\cdot\bv\times\bOmega=(\varepsilon\rho\dot\theta)\bn\cdot\bv\times\bb=
-(\varepsilon\rho\dot\theta)\bv\cdot\bc=-(\varepsilon\rho\dot\theta)w\,.
\]
Then, replacing in \eqref{LittleLagr} and averaging over the gyrophase yields the averaged Lagrangian
\beq\label{LLagr}
 \langle{L}\rangle=\big(u\bb+\varepsilon^{-1}\bA(\bX)\big)\cdot\dot{\bf X}
+\Big(\frac{1}2\varepsilon^{-1}\rho_{L\,} B(\bX)-w\Big)\rho_L\dot\theta
-\frac{1}2\big(u^2+w^2+n^2\big)
\eeq
This Lagrangian differs from that presented by Littlejohn (see equation (29) in \cite{Littlejohn}), since it was obtained without using any extra information on the relation between gyroradius and the perpendicular velocity. Indeed, the relation
 \[
 \rho_L=\frac{\varepsilon w}{B(\bX)}
 \]
is obtained from the Lagrangian above only after taking variations in $\rho_L$. We emphasize that this is the main advantage of the present approach: no previous knowledge on the gyrophase  and the gyroradius is assumed, except their fast rotational dynamics. Variations in $w$ return
 $\dot\theta=-\varepsilon^{-1}B(\bX)$
as in the previous Section, while variations in $n$ simply yield $n=0$ (zero velocity along the gyroradius). Again, notice that this last result is obtained naturally from Hamilton's principle, because no assumption on $n$ was ever invoked. 
In addition, conservation of the magnetic moment in the form $\rho_L^2B(\bX)=const.$ is obtained by taking variations in $\theta$, while Littlejohn's equations \eqref{LGC} (with $\nabla\times\bb=0$ and in the absence of an electric field, i.e. $\bE=-\partial_t\bA=0$) are obtained by taking variations in $u$ and $\bX$.
 Notice that, although \eqref{LLagr} differs from Littlejohn's Lagrangian, the latter is recovered by replacing $n=0$ and $\rho_L={\varepsilon w}/{B(\bX)}$. Indeed, this yields
 \[
 \langle{L}\rangle=\Big(\varepsilon^{-1}\bA(\bX)+u\bb\Big)\cdot\dot{\bf X}
-\frac{\varepsilon w^2}{2 B(\bX)}\dot\theta-\frac{1}2\big(u^2+w^2\big)
\,,
 \]
 which is precisely {Littlejohn's Lagrangian}  upon changing $\theta$ by $-\theta$.

While these last two Sections have presented the simplest nontrivial case of a magnetic field with constant direction, we shall see that inhomogeneities of the magnetic field leave this construction completely unaffected with the only result that the the Lagrangian \eqref{LLagr} will now involve an inhomogeneous magnetic field $\bb(\bX)$, thereby producing extra terms in the equations of motion. This is explained in more detail in the next Section.

\subsection{Inhomogeneous static magnetic field \label{InhomSt}}
In this section, we shall show that the case of magnetic field with varying direction (i.e., $\bb=\bb(x)$) does not involve any modifications to the treatment above. Indeed, Northrop's Lagrangian \eqref{NLagrangian}  and Littlejohn's Lagrangian (in the form \eqref{LLagr}) are left unaffected by the inhomogeneities of the magnetic field. The terms arising from these inhomogeneities are either higher-order or they average to zero.

Here, we first approach the problem in Northrop's picture. Upon recalling \eqref{connection1}, we shall expand the  Lagrangian \eqref{Lagr-config} by using the relation
\begin{align*}
\dot{\brho}(\bX,t)
=&\ 
\dot\rho(t)\bn(\bX,t)+\varepsilon^{-1}\dot\Theta(t)\bb\times\brho(\bX,t)-\dot{X}^i\bgamma_i(\bX)\times\brho(\bX,t)
\,,
\label{gyrorad-exp2}
\end{align*}
which modifies \eqref{gyrorad-exp} in order to account for the inhomogeneities of the magnetic field. We expand the kinetic energy in \eqref{Lagr-config} as
\begin{align*}
\frac{\varepsilon}2\|\dot{\bf r}\|^2=&\,\frac{\varepsilon}2\big\|\dot\bX+\varepsilon\big( \dot\rho\boldsymbol{a}-\dot{X}^i\bgamma_i\times\brho\big)+ \dot\Theta\bb\times\boldsymbol{\rho}\big\|^2
\\
=&\,
\frac{\varepsilon}2\,\|\dot\bX\|^2+ \frac{\varepsilon}2\rho^2\dot\Theta^2+\dot\Theta\dot\bX\cdot \bb\times\boldsymbol{\rho}+O(\varepsilon)
\,,
\end{align*}
while expanding the minimal coupling term $\dot{\bf r}\cdot\bA$ in \eqref{Lagr-config} yields
\begin{multline*}
\left(\bA+\varepsilon\boldsymbol{\rho}\cdot\nabla\bA\right)\cdot\big[\dot{\bf X}+\varepsilon\big( \dot\rho\boldsymbol{a}-\dot{X}^i\bgamma_i\times\brho\big)+ \dot\Theta\bb\times\boldsymbol{\rho}\big]
\\
=
\bA\cdot\big[\dot{\bf X}+\dot\Theta\bb\times\brho+\varepsilon\big( \dot\rho\boldsymbol{a}-\dot{X}^i\bgamma_i\times\brho\big)\big]
+
\varepsilon\brho\cdot\nabla\bA\cdot\big[\dot{\bf X}+ \dot\Theta\bb\times\brho\big]+O(\varepsilon)
\,.
\end{multline*}
Upon comparing with the results in Appendix \ref{NorthApp}, this calculation shows that expanding the Lagrangian \eqref{Lagr-config} to first order leads to the emergence of the following extra term in \eqref{NorthLagr}:
\beq\label{extraterm}
\varepsilon\brho\cdot\dot{X}^i\bgamma_i\times\bA
\,,
\eeq
which however vanishes upon averaging because of \eqref{averages}. Thus, the presence of inhomogeneities in the direction of the magnetic field does not affect the expression of Northrop's Lagrangian \eqref{NLagrangian}, which in turn yields  the guiding center motion \eqref{NGC} with a static magnetic field and in the absence of an electric field, i.e. $\bE=-\partial_t\bA=0$.

As it is straightforward to verify, a completely analogue situation holds in Littlejohn's phase-space picture. Indeed, expanding the phase-space Lagrangian \eqref{phspLagr} upon retaining the inhomogeneities of the magnetic field produces precisely the same extra term \eqref{extraterm} in \eqref{LittleLagr}. However, this term is eliminated by averaging and therefore the final Lagrangian \eqref{LLagr} for guiding center motion is left unaffected. Nevertheless, variations of the term $u\bb\cdot\dot{\bf X}$ in \eqref{LLagr} lead to non-trivial terms  (since $\nabla\times\bb\neq0$) in the resulting dynamics, which is given by  \eqref{GCEQ} (upon specializing to the static case $\partial_t\bb=0$).

\section{External time-dependent magnetic field\label{DynMagnF}}

In presence of an external dynamic magnetic field, the previous approach to guiding center motion needs to be modified in order to split \emph{fast rotations} around the magnetic field and  \emph{slow rotations} of the magnetic field. This is equivalent to a slow-fast timescale decomposition. In order to generalize the previous approach to dynamic magnetic fields, we start by writing the gyrodirector evolution as
\[
\bn(\bx,t)=\Lambda(\bx,t)\bn_0(\bx)
\,.
\]
Here, the rotation matrix $\Lambda(\bx,t)$ incorporates the overall rotational state of the gyroradius, including slow rotations of the magnetic field. These rotations can be naturally incorporated by writing the $\bb$ evolution as 
\beq\label{magneticevol}
\bb(\bx,t)=\chi(\bx,t)\bb_0(\bx)
\,,
\eeq
where $\chi(\bx,t)$ is the rotational state of the magnetic field. Taking the time derivative one has
\[
\bb(\bx,t)=\bnu(\bx,t)\times\bb(\bx,t)
\,,
\]
where $\widehat{\nu}=(\partial_t\chi)\chi^{-1}$. We ask that $\chi$ does not involve any rotation around the magnetic field: this amounts to a gauge fixing, that is obtained by $\bnu\cdot\bb=0$. Therefore, the only choice is
\beq\label{freq2}
\bnu(\bx,t)=\bb(\bx,t)\times\partial_t\bb(\bx,t)
\,,
\eeq
so that $\widehat{\nu}=\big[_\text{\tiny$\,$}\widehat{b},\partial_t\widehat{b}_\text{\tiny$\,$}\big]$, in commutator notation. 
In turn, this determines $\chi$ by solving the reconstruction equation $\partial_t\chi=\big[_\text{\tiny$\,$}\widehat{b},\partial_t\widehat{b}_\text{\tiny$\,$}\big]\chi$.

\subsection{Splitting slow and fast rotations}
At this point, one must find the relation between the rotational state $\Lambda$ of the gyroradius and that of the magnetic field $\chi$. The initial gyroradius $\bn_0$ may reach its current position $\bn$ in two possible ways: the first is $\Lambda=\mathcal{R}(\theta,\bb)\chi$, while the second is $\Lambda=\chi\mathcal{R}(\theta,\bb)$. The first way corresponds to first rotating $\bb$ and then rotating around it, while the second one involves a rotation around the magnetic field followed by a rotation of the magnetic field itself. However, only one choice is allowed because $\bn$ and $\bb$ are required to be orthogonal at all times. Since one has
\[
\bb\cdot\chi\mathcal{R}(\theta,\bb)\bn_0=\bb_0\cdot\mathcal{R}(\theta,\bb)\bn_0\neq\bb_0\cdot\bn_0\,,
\]
the only possible choice is 
\[
\Lambda=\mathcal{R}(\theta,\bb)\chi
\,.
\]
Indeed one easily verifies that $\bb\cdot\mathcal{R}(\theta,\bb)\chi\bn_0=\bb\cdot\chi\bn_0=\bb_0\cdot\bn_0$. Therefore, the correct evolution law for the gyroradius reads
\beq\label{gyroevolution2}
\bn(\bx,t)=\mathcal{R}(\theta(t),\bb(\bx,t))\chi(\bx,t)\bn_0(\bx)
\,.
\eeq
Equivalently, one can write $\bn(\bx,t)=\mathcal{R}(\theta(t),\bb(\bx,t))\mathbf{e}_1(\bx,t)$, where we have defined $\mathbf{e}_1(\bx,t)=\chi(\bx,t)\bn_0(\bx)$. With this last definition, multiplying by $\rho(t)$ on both sides and replacing $\theta=-\vartheta$, equation \eqref{gyrodirector} yields exactly the first in \eqref{Northrop}  (see also equations (1.2) in \cite{Littlejohn}). Equation \eqref{gyroevolution2} means that the fast rotations around the magnetic field act on the slow rotations of the magnetic field itself to give the overall motion. Notice that the gauge fixing $\bnu\cdot\bb=0$ in \eqref{freq2} implies $\bb\times\mathbf{e}_1\cdot\partial_t\mathbf{e}_1=0$, since $(\bnu\times\mathbf{e}_1)\cdot(\bb\times\mathbf{e}_1)=\bnu\cdot\bb=0$ (the same relations hold upon replacing $\mathbf{e}_1$ by ${\mathbf{e}_2}=\bb\times{\mathbf{e}_1}$).

Still, the evolution law \eqref{gyroevolution2} has the complication that $\bb$  involves $\chi$ via \eqref{magneticevol}. This difficulty can be overcome by noticing that 
\[
\mathcal{R}(\theta,\bb)=\exp(\theta\widehat{\bb}_\text{\tiny$\,$})= \exp(\theta\chi\widehat{\bb}_0\chi^{-1})=\chi \exp(\theta\widehat{\bb}_0)\chi^{-1}= \chi \mathcal{R}(\theta,\bb_0)\chi^{-1}
\,,
\]
where we have used the relation $\widehat{b}=\chi\widehat{\bb}_0\chi^{-1}$, which can be verified explicitly upon using  \eqref{magneticevol}.
Consequently, \eqref{gyroevolution2} can be rewritten in a simpler way as
\beq\label{gyroevolution3}
\bn(\bx,t)=\chi(\bx,t)\mathcal{R}(\theta(t),\bb_0(\bx))\bn_0(\bx)
\,,
\eeq
so that the slow timescales are all encoded in the rotation $\chi$, while the matrix $\mathcal{R}(\theta,\bb_0)=e^{\theta\widehat{b}_0}$ only involves the fast timescales. In what follows, we shall simply write $\mathcal{R}$ to denote $\mathcal{R}(\theta,\bb_0)$. Then, taking the time derivative of \eqref{gyroevolution3} gives the equation of motion
\beq\label{a-dyn}
\partial_t\bn=\bnu\times\bn+\chi\dot{\mathcal{R}}\mathcal{R}^{-1} \mathcal{R}\bn_0=\bnu\times\bn+\dot\theta\chi\widehat{b}_0 \chi^{-1}\chi \mathcal{R}\bn_0=(\bb\times\partial_t\bb+\dot\theta\bb)\times\bn
\,,
\eeq
where we have used $\dot{\mathcal{R}}\mathcal{R}^{-1}=\dot\theta_\text{\tiny$\,$}\widehat{b}_0$ and $\widehat{b}_\text{\tiny$\,$}\bn=\bb\times\bn$, as well as \eqref{freq2}  and \eqref{gyroevolution3}. Also, computing the gradients of \eqref{magneticevol} 
yields
\[
\nabla_i\bb=-\boldsymbol\kappa_i\times\bb
\,,\qquad
\widehat\kappa_i=\chi\nabla_i\chi^{-1}+\chi\widehat{\kappa}_{0\,i}\chi^{-1}
\,,
\]
where we have chosen $\bkappa_0$ such that $(\nabla_i\bn_0,\nabla_i\bb_0)=-(\bkappa_{0\,i}\times\bn_0,\bkappa_{0\,i}\times\bb_0)$. The wryness tensor $\bkappa$ can be expressed in analogy with \eqref{wrynessbiaxial} as $\bkappa_{i}=\nabla_i \bb\times \bb+(\nabla_i \mathbf{e}_1\cdot \mathbf{e}_1\times \bb)\bb$. Its dynamics is given by (see  Section \ref{nematics}) $\partial_t\bkappa_i=-\nabla_i\bnu-\bkappa_i\times\bnu$, where we recall equation \eqref{freq2}. This equation of motion allows to express $\bkappa$ only in terms of the unit magnetic field: upon dotting by $\bb$, one has $\partial_t(\bb\cdot\bkappa_i)=-\nabla(\bb\times\partial_t\bb)\cdot\bb$ and this can be further expanded to give
\beq\label{bdotkappa}
\bb\cdot\bkappa_i=\nabla_i \mathbf{e}_1\cdot \mathbf{e}_1\times \bb=\bb\cdot\bkappa_{0\,i}+\int^t\!\bb\cdot(\partial_\tau\bb)\times(\nabla_i\bb)\,\de\tau
\,.
\eeq
Eventually, the final expression of $\bkappa$ reads
\[
\bkappa_i=\nabla_i \bb\times \bb+\left[\nabla_i \bn_0\cdot \boldsymbol{c}_0+\int^t\!\bb\cdot(\partial_\tau\bb)\times(\nabla_i\bb)\,\de\tau\right]\bb
\,,
\]
where we recall $\mathbf{e}_1(\bx,0)=\bn_0(\bx)$ and we have defined $\boldsymbol{c}_0(\bx)=\bn_0(\bx)\times\bb_0(\bx)$ (see previous sections).
The tensor $\bkappa$ is the wryness tensor corresponding to the slow rotations $\chi$ of the magnetic field.

\subsection{The gyro-wryness tensor}
In the case of a time-dependent magnetic field, the gyro-wryness tensor appears by taking gradients of equation \eqref{gyroevolution3}. Indeed, one has
\beq\label{New-gyrowryn}
\nabla_i\bn=-\boldsymbol\Gamma_i\times\bn
\,,\qquad
\nabla_i\bb=-\boldsymbol\Gamma_i\times\bb
\,,\qquad
\widehat\Gamma_i=(\chi \mathcal{R})\nabla_i(\chi \mathcal{R})^{-1}+\chi \mathcal{R}\widehat{\kappa}_{0\,i}(\chi \mathcal{R})^{-1},
\eeq
where the second follows from $\bb=\chi\bb_0=\chi\mathcal{R}\bb_0$. Expanding out the definition of $\widehat{\Gamma}$ gives
\[
\widehat\Gamma_i=\chi\nabla_i\chi^{-1}+\chi\left[\mathcal{R}\nabla_i \mathcal{R}^{-1}+ \mathcal{R}\widehat{\kappa}_0\mathcal{R}^{-1}\right]\chi^{-1}=\widehat{\kappa}_i+\chi(\widehat{\gamma}_{i}-\widehat{\kappa}_{0\,i})\chi^{-1}
\,,
\]
where $\widehat{\gamma}_i=\mathcal{R}\nabla_i \mathcal{R}^{-1}+ \mathcal{R}\widehat{\kappa}_{0\,i}\mathcal{R}^{-1}$ is the wryness tensor corresponding to fast rotations $\mathcal{R}$ around the magnetic field. We remark that $\widehat{\gamma}_i(\bx,0)=\widehat{\kappa}_{0\,i}(\bx)$ by construction. Since $\widehat{\gamma}_i(\bx,t)$ is defined exactly as the gyro-wryness tensor in Section \ref{sec:gyrodirector}, we have $\bgamma_i=\nabla_i{\bb}_0\times{\bb}_0+(\bb_0\cdot\bgamma_i){\bb}_0$ and
\[
\bb_0\cdot\bgamma_i=\bb_0\cdot\bkappa_{0\,i}=\nabla_i \bn_0\cdot \boldsymbol{c}_0
\]
At this point, it is convenient to write $\widehat\Gamma_i$ in vector notation as
\beq
\boldsymbol{\Gamma}_i=\bkappa_i+\chi(\bgamma_{i}-\bkappa_{0\,i})
\,.
\label{NewGyrowrynExp}
\eeq
Indeed, since $\bb\cdot\chi(\bgamma_{i}-\bkappa_{0\,i})=\bb_0\cdot(\bgamma_{i}-\bkappa_{0\,i})=0$, we have $\bb\cdot\boldsymbol{\Gamma}_i=\bb\cdot\bkappa_i$ and
\beq\label{R-evolution}
\bb\cdot\boldsymbol{\Gamma}_i=-\nabla_i\mathbf{e}_1\cdot\mathbf{e}_2=\nabla_i \bn_0\cdot \boldsymbol{c}_0+\!\int^t\!\bb\cdot(\partial_\tau\bb)\times(\nabla_i\bb)\,\de\tau=-R_i
\eeq
It is interesting to compare our expression for $\mathbf{R}$ with its equation of motion, as it was given by Littlejohn (see equation (4.3) in \cite{Littlejohn88}), which reads
\[
\frac{\partial\mathbf{R}}{\partial t}=-\nabla\!\left(\frac{\partial\mathbf{e}_1}{\partial t}\cdot\bb\times\mathbf{e}_1\right)-\nabla\bb\cdot\!\left(\bb\times\frac{\partial\bb}{\partial t}\right)
\]
where we recall that $\partial_t\mathbf{e}_1=\bnu\times\mathbf{e}_1$. 
As previously observed, our gauge choice $\bnu\cdot\bb=0$ enforces the first term on the right hand side above to vanish, thereby recovering \eqref{bdotkappa}. In analogy with  electromagnetism, we call the gauge ${\partial_t\mathbf{e}_1}\cdot\bb\times\mathbf{e}_1=0$ the \emph{Weyl gauge}: this particular gauge allows to express $\mathbf{R}$ in terms of the magnetic field (up to an initial condition), as explicitly shown in \eqref{R-evolution}. The Weyl gauge emerges here as the natural gauge for Littlejohn's $\mathbf{R}-$vector: again, this result is achieved here by exploiting the analogies between guiding center motion and liquid crystal dynamics. Notice that in the case of an initial magnetic field with constant direction, one can set $\nabla\bn_0=0$, thereby yielding 
\[
\mathbf{R}=-\!\int^t\!\bb\cdot(\partial_\tau\bb)\times(\nabla\bb)\,\de\tau
\,.
\]
However, as pointed out by Littlejohn \cite{Littlejohn88}, in the general case  there is no possible nontrivial choice of $\bn_0$ such that $\nabla_i\bn_0\cdot \bb_0\times\bn_0=0$ for all $i$. See also \cite{BuQi} for more recent work on this point. 
The relations obtained here can be used in higher order theory to replace the term $\varepsilon^2\mu\mathbf{R}\cdot\dot\bX$, which appears in Littlejohn's Lagrangian upon retaining higher order terms \cite{Littlejohn}. 

The next Section shows how the present construction is applied directly to  the theory of guiding center motion.

\subsection{Guiding center dynamics}

By proceeding analogously to the treatment in Section \ref{InhomSt} for inhomogeneous magnetic fields, we shall show here that the dynamics of the external magnetic field does not affect the final expressions of Northrop's Lagrangian \eqref{NLagrangian}  and Littlejohn's Lagrangian (in the form \eqref{LLagr}). We begin by writing 
\begin{align}\nonumber
\dot{\brho}=&\,\dot{\rho}\bn+\rho\partial_t\bn+\rho \dot{X}^i\boldsymbol{\Gamma}_i\times\bn
\\
=&\,
\dot{\rho}\bn+\rho (\dot{X}^i\boldsymbol{\Gamma}_i+\bb\times\partial_t\bb)\times\bn+\varepsilon^{-1}\rho\dot\Theta\bb\times\bn
\,,
\end{align}
where we recalled \eqref{a-dyn}, \eqref{New-gyrowryn} and \eqref{NewGyrowrynExp}. Upon expanding the Lagrangians \eqref{Lagr-config} and \eqref{phspLagr} to first order, we verify that the minimal coupling term $\dot{\bf r}\cdot \bA$ produces the following extra term in \eqref{NorthLagr} and \eqref{LittleLagr}
\[
-\varepsilon\brho\cdot\bA\times(\dot{X}^i\boldsymbol{\Gamma}_i+\bb\times\partial_t\bb)
\,,
\]
which emerges from the inhomogeneities of a time dependent magnetic field. However, similarly to the situation in Section \ref{InhomSt}, the term above averages to zero, thereby leaving the expressions of Northrop's Lagrangian \eqref{NLagrangian}  and Littlejohn's Lagrangian (in the form \eqref{LLagr}) unaffected. Then, Northrop's guiding center equation \eqref{NGC} now accounts for a dynamic magnetic field $\bB(\bx,t)$ and electric field $\bE(\bx,t)=-\partial_t\bA(\bx,t)$, while Littlejohn's equations \eqref{GCEQ} also exhibit a nonzero $\partial_t\bb(\bX,t)$ (which arises from taking variations of the term $u\bb\cdot\dot\bX$).

\section{Summary}

This work has exploited geometric techniques in liquid crystal theories to shed new light on guiding center motion. After combining the WKB ordering method with the rotational ansatz for the gyroradius, averaging the Lagrangian yields a new variational principle, which does not assume any prior knowledge on the gyroradius or the gyrophase (except their fast rotational motion). This construction was presented for lowest order guiding center theory and its possible extensions to higher order dynamics is out of the scope of this paper. Higher-order guiding center theory still greatly benefits from the use of sophisticated techniques such as Lie transforms and Hamiltonian perturbation theory \cite{CaLi,Littlejohn1981,Littlejohn1982}.

In this work, new Lagrangian structures were obtained for guiding center motion in both Northrop and Littlejohn's pictures for both static and time-dependent magnetic fields. In this new approach, the slow rotations of the magnetic field were separated from the fast rotations around the magnetic field, thereby generating new explicit expressions for the gyrowryness tensor. As a consequence, the Weyl gauge emerged as a natural gauge that enables writing Littlejohn's vector $\bf R$ uniquely in terms of the magnetic field (up to arbitrary initial conditions).

Due to its flexibility, this construction can be fruitfully used for modeling purposes, for example in the formulation of nonlinear hybrid kinetic-fluid systems \cite{HoTr2011,Tronci2010,TrTaCaMo}. In this context, a certain class of nonlinear hybrid MHD models (pressure-coupling scheme) involves writing the kinetic theory for energetic particle in the Lagrangian frame of the MHD bulk fluid \cite{HoTr2011}. While guiding-center motion has been written in moving Eulerian frames \cite{Brizard}, its formulation with respect to Lagrangian fluid frames can be approached by the techniques presented in this paper and this is part of ongoing work.

Another interesting direction emerges from the appearance of the gyro-wryness tensor for dynamic magnetic fields: one could even ask if the construction presented here can be used to study magnetic defects, that is points (or sets of points) where the unit vector $\bb$ is not defined. Indeed, it was shown in \cite{BuQi} that in some cases the gyrophase may not  be defined globally. For example, if guiding-center dynamics is coupled to Maxwell equations \cite{AJBCT,Pfirsch,PfMo}, magnetic reconnection can occur thereby yielding an $X$-point where $\bb$ is undefined. It is an interesting question whether the description of the $X$-point in magnetic reconnection can be approached by the theoretical construction presented in this work.

\medskip

\paragraph{Acknoweledgments.} The author is greatly indebted to Paul Skerritt for his keen remarks and geometric insight, which were determinant to the development of this work.  In addition, the author is grateful to Alain Brizard, Joshua Burby, Enrico Camporeale, Cristel Chandre, Alexander Close, Fran\c{c}ois Demures, Fran\c{c}ois Gay-Balmaz, Darryl Holm, Philip Morrison, and Emanuele Tassi for their valuable comments on these results. Financial support by the Leverhulme Trust Research Project Grant 2014-112 is also acknowledged.

\bigskip

\appendix

\section{Appendix}

\subsection{Ordering for Northrop's Lagrangian\label{NorthApp}}
In this appendix, we derive the Lagrangian \eqref{NorthLagr}, upon retaining the assumption of a magnetic field with constant direction. 

The treatment is similar to that in Section III.C of \cite{CaBr}.
Consider the Lagrangian \eqref{Lagr-config} for a charged particle in a magnetic field 
and recall the relations \eqref{gyroexpansion}, \eqref{gyroradius1}, \eqref{gyrodirector}, \eqref{gyrorad-exp}, and \eqref{newangle}. 
Then, we compute
\beq
\dot{\bf r}=\dot{\bf X}+\varepsilon \dot\rho\boldsymbol{a}+ \rho\dot\Theta\bb\times\boldsymbol{a}
\label{usefulformula2}
\eeq
and expand the kinetic energy in \eqref{NorthLagr} to first order:
\begin{align*}
\frac{\varepsilon}2\|\dot{\bf r}\|^2=&\,\frac{\varepsilon}2\big\|\dot\bX+\varepsilon \dot\rho\boldsymbol{a}+ \rho\dot\Theta\bb\times\boldsymbol{a}\big\|^2
\\
=&\,
\frac{\varepsilon}2\,\|\dot\bX\|^2+ \frac{\varepsilon}2\rho^2\dot\Theta^2+\rho\dot\Theta\,\dot\bX\cdot \bb\times\boldsymbol{a}+O(\varepsilon)
\,.
\end{align*}
Also, we expand $\bA({\bf r})=\bA(\bX)+\varepsilon\brho\cdot\nabla\bA(\bX)$ and compute
\begin{align*}
\left(\bA+\varepsilon\rho\boldsymbol{a}\cdot\nabla\bA\right)\cdot\big(\dot{\bf X}+\varepsilon \dot\rho\boldsymbol{a}+ \rho\dot\Theta\bb\times\boldsymbol{a}\big)
&=
\big(\bA+\varepsilon\rho\boldsymbol{a}\cdot\nabla\bA\big)\cdot\dot{\bf X}+\varepsilon \dot\rho\bA\cdot\boldsymbol{a}
\\
&\quad+\varepsilon\rho\dot\Theta\big(\bA+\rho\boldsymbol{a}\cdot\nabla\bA\big)\cdot\bb\times\boldsymbol{a}+O(\varepsilon)
\\
&=
\bA\cdot\big(\dot{\bf X}+\rho\dot\Theta\bb\times\boldsymbol{a}\big)+\varepsilon \dot\rho\bA\cdot\boldsymbol{a}+\varepsilon\rho\boldsymbol{a}\cdot\nabla\bA\cdot\dot{\bf X}\
\\
&\quad+\varepsilon\rho^2\dot\Theta\boldsymbol{a}\cdot\nabla\bA\cdot\bb\times\boldsymbol{a}+O(\varepsilon)
\,.
\end{align*}
Upon focusing on the last term,  we observe
\begin{align*}
\varepsilon\dot\Theta\boldsymbol{a}\cdot\nabla\bA\cdot\bb\times\boldsymbol{a}
=&\,
\varepsilon^2\boldsymbol{a}\cdot\nabla\bA\cdot\dot{\boldsymbol{a}}
\\
=&\,
\frac12\varepsilon^2\boldsymbol{a}\cdot[\nabla\bA-\nabla\bA^T]\cdot\dot{\boldsymbol{a}}
+
\frac12\varepsilon^2\boldsymbol{a}\cdot[\nabla\bA+\nabla\bA^T]\cdot\dot{\boldsymbol{a}}
\\
=&\,
\frac12\varepsilon^2\boldsymbol{a}\cdot\dot{\boldsymbol{a}}\times\bB
+
\frac12\varepsilon^2\!\left(\frac{\de}{\de t}(\boldsymbol{a}\cdot\nabla\bA\cdot\boldsymbol{a})-\boldsymbol{a}\cdot\frac{\de\nabla\bA}{\de t}\cdot\boldsymbol{a}\right)
\end{align*}
and 
\[
\frac12\varepsilon^2\rho^2\frac{\de}{\de t}(\boldsymbol{a}\cdot\nabla\bA\cdot\boldsymbol{a})
=
\frac12\varepsilon^2\frac{\de}{\de t}(\rho^2\boldsymbol{a}\cdot\nabla\bA\cdot\boldsymbol{a})
-
\varepsilon^2\rho\dot{\rho}\boldsymbol{a}\cdot\nabla\bA\cdot\boldsymbol{a}\,.
\]
Then, upon combining the results above and by retaining only the first order terms, we obtain (see Section III.C in \cite{CaBr})
\beq
\dot{\bf r}\cdot\bA({\bf r},t)
=
\bA\cdot\dot{\bf X}+\varepsilon\rho\boldsymbol{a}\cdot\dot\bX\times\bB
+\frac\varepsilon2\rho^2B\dot\Theta+O(\varepsilon)+\frac{\de \Phi}{\de t}
\,,
\label{usefulformula}
\eeq
where $\Phi(\bX,\rho,\Theta)$ is some function that can be discarded since we recall that exact time derivatives in the Lagrangian are irrelevant to the dynamics.

Therefore, upon dividing by $\varepsilon$ and by averaging over $\theta=\varepsilon^{-1}\Theta$ using \eqref{averages}, the Lagrangian becomes
\begin{align*}
\langle{L}\rangle=&\,
\frac12\|\dot\bX\|^2+\varepsilon^{-1}
\bA\cdot\dot{\bf X}+ \frac12 \rho^2\dot\Theta\big(\dot\Theta+ B\big),
\end{align*}
which coincides with \eqref{NorthLagr} by recalling \eqref{gyroradiusconversion}.

\subsection{Ordering for the phase-space Lagrangian\label{LittleApp}}
In this Appendix, we derive the Lagrangian \eqref{LittleLagr}, upon retaining the assumption of a magnetic field with constant direction. The treatment makes use of the relations \eqref{usefulformula2} and \eqref{usefulformula}.

Consider the phase-space Lagrangian \eqref{phspLagr} for a charged particle moving in a magnetic field 
and recall the relations \eqref{gyroexpansion}, \eqref{gyroradius1}, \eqref{gyrodirector}, \eqref{gyrorad-exp}, and \eqref{newangle}. Since the term $\dot{\bf r}\cdot\bA({\bf r},t)$ has been already treated in the previous section, we consider
\[
\epsilon\bv\cdot\dot{\bf r}
=
\epsilon\bv\cdot\big(\dot{\bf X}+ \rho\dot\zeta\bb\times\boldsymbol{a}\big)+O(\epsilon)
\,.
\]
Then, by following the same steps as those preceding \eqref{usefulformula}, the phase-space Lagrangian becomes
\begin{align*}
L=&\,\Big(\bv+\epsilon^{-1}\bA\Big)\cdot\dot{\bf X}+\epsilon\rho\bv\cdot\dot{\boldsymbol{a}}+\rho\boldsymbol{a}\cdot\dot\bX\times\bB
+\frac12\epsilon\rho^2\boldsymbol{a}\cdot\dot{\boldsymbol{a}}\times\bB-\frac{1}2\|\bv\|^2
\\
=&\,
\Big(\bv+\epsilon^{-1}\bA\Big)\cdot\dot{\bf X}+\rho\dot\zeta\bv\cdot\bb\times{\boldsymbol{a}}+\rho\boldsymbol{a}\cdot\dot\bX\times\bB
+\frac12\rho^2B\dot\zeta-\frac{1}2\|\bv\|^2
\,,
\end{align*}
which can be rewritten as \eqref{LittleLagr} by recalling \eqref{magneticmoment} as well as the definition $\bOmega=\dot\Theta\bb$.

\end{document}